\documentclass{appolbaXv}
\usepackage{graphicx}
\usepackage{subfigure}
\newcommand{\eq}{\begin{equation}}
\newcommand{\eqx}{\end{equation}}
\newcommand{\eqn}{\begin{eqnarray}}
\newcommand{\eqnx}{\end{eqnarray}}
\newcommand{\bi}{\begin{itemize}}
\newcommand{\ei}{\end{itemize}}
\newcommand{\nn}{\nonumber}
\newcommand{\ra}{\rangle}
\newcommand{\la}{\langle}
\newcommand{\bz}{\bar{z}}

\begin{document}
\title{\bf Beyond complex Langevin equations II: a positive representation of Feynman path integrals directly in\\ the Minkowski time
}
\author{Jacek Wosiek\thanks{e-mail: wosiek@th.if.uj.edu.pl}
\address{M. Smoluchowski Institute of Physics, Jagellonian University\\
\L ojasiewicza St. 11, 30-348 Krakow}
}
\maketitle
\begin{abstract}
Recently found positive representation for an arbitrary complex, gaussian weight is used to construct a statistical formulation of gaussian path integrals directly in the Minkowski time. The positivity of Minkowski weights is achieved by doubling the number of real variables. The continuum limit of the new representation exists only if some of the additional couplings tend to infinity and are tuned in a specific way. The construction is then successfully applied to three quantum mechanical examples including a particle in a constant magnetic field -- a simplest prototype of a Wilson line. Further generalizations are shortly discussed and an intriguing 
interpretation of new variables is alluded to.
\end{abstract}
  
\section{Introduction and a single integral}
Stochastic quantization \cite{PW,DH} based on complex Langevin equations \cite{P,Kl} has attracted again a new wave of interest. This was caused by reported lately progress in simulating lattice QCD at finite chemical potential \cite{S1,S2}.  At the same time the old problems \cite{AY,AFP,HW}, the approach suffered with, resurfaced again \cite{Ph,Bl} only to emphasize the difficulty with theoretical foundations of the method \cite{S3,S4}.
 
Recently a positive representation, equivalent to the complex gaussian distribution in the complex Langevin approach,  
was derived \cite{Ja}. The problem is not new and its classic, by now, solution is known for a long time \cite{AY}. The novelty of the present
result is that it provides a positive representation for an arbitrary complex value of the inverse dispersion parameter $\sigma$, while the original one
applies for ${\cal R}e\; \sigma > 0$ only. In particular the new solution works also for purely imaginary $\sigma$. This opens a possibility of a positive representation 
for Feynman path integrals directly in the Minkowski time -- the quest which still awaits its resolution. 

In this article it is shown that indeed such an extension is possible. It is constructed and applied to few quantum mechanical textbook cases.
Noteworthy, the construction covers also the path integral description of a particle in a constant magnetic field -- a problem which does not have a positive representation after the Wick rotation.

In 2002 Weingarten \cite{We} has addressed analogous question in more general terms and has proved that the positive densities actually exist
for a wide class of complex probabilities. Nevertheless no practical construction of such distributions was attempted even for the gaussian case  (see however \cite{Sa}).
Moreover, the continuum limit was explicitly not discussed. As will be seen below, existence of the continuum limit
plays an essential role in the present solution. 

To begin with, we recall the idea of Ref.\cite{Ja} for a single integral.  In general, quantum averages result from weighting observables
with complex functions $\rho(x)\equiv e^{-S(x)}$, rather than with positive probabilities. The complex Langevin approach can in principle
address this difficulty by replacing a complex average with the statistical average over the complex stochastic process determined by a complex action $S(x)$
\eqn
\frac{\int f(x) e^{-S(x)} dx}{\int e^{-S(x)} dx } = \frac{\int \int f(x+i y) P(x,y) dx dy }{\int \int P(x,y) dx dy}, 
\label{B1}
\eqnx
with $P(x,y)$ being the distribution of the above process at large Langevin time. While this idea works well and has been proven
for real actions, it encounters theoretical doubts and practical difficulties in the complex case.  Instead we have  constructed $P(x,y)$  directly using (\ref{B1}) as a starting point and avoiding any reference to stochastic processes and associated Fokker-Planck equations. The derivation works as follows. Introduce two independent, complex variables
\eqn
\;\;\;z=x+iy,\;\;\;\bz=x-iy,     \label{xy}
\eqnx
and rewrite (\ref{B1}) as
\eqn
\frac{\int_R f(x) 
\rho(x) dx}{\int_R \rho(x) dx }=
 \frac{\int_{\Gamma_z} f(z) \rho(z) dz}{\int_{\Gamma_z} \rho(z) dz} =\nn \\ \frac{\int_{\Gamma_z} \int_{\Gamma_{\bz}} f(z) P(z,\bz) dz d\bz }{\int_{\Gamma_z} \int_{\Gamma_{\bz}} P(z,\bz) dz d\bz}
 = \frac{\int_{R^2} f(x+iy) P(x,y) dx dy }{\int_{R^2} P(x,y) dx dy}. \label{B3}
\eqnx
Contours $\Gamma_z$ and $\Gamma_{\bz}$ are such that the integrals exists. Above equations will be satisfied provided we find $P(z,\bz)$ such that
\eqn
\rho(z)=\int_{\Gamma_{\bz}} P(z,\bar{z}) d \bar{z}.  \label{Pro}
\eqnx
This is the key relation of the new approach. On one side it provides a simple connection between a complex weight $\rho$ and $P$, while on the other leaves us a freedom 
to satisfy positivity and normalizability of $P(z,\bz)$ restricted to $R^2$ by (\ref{xy}).  For more details, also in a simple quartic case, see \cite{Ja}. 

For the gaussian action we take 
\eqn
P(z,\bz)=\frac{i}{2}\exp{\left(-(a^* z^2 + 2 b z \bz + a \bz^2)\right)},\;\;\;\;\; a=\alpha+i\beta,\; b=b^*, \label{Pzz}
\eqnx
or in terms of $x$ and $y$,
\eqn
P(x,y)=\exp{\left(-2\left((b+\alpha) x^2 + 2\beta x y + (b-\alpha) y^2\right)\right)}, \label{Pxy}
\eqnx
which is positive and normalizable for $|a| < b$. Corresponding complex density follows from (\ref{Pro})
\eqn
\rho(z)=\int_{\Gamma_{\bz}} P(z,\bz) d \bz = 
  \frac{1}{2}\sqrt{\frac{\pi}{-a}}\exp{\left( - \sigma z^2\right)},\;\;\;\sigma=\frac{|a|^2-b^2}{a}, \nn 
\eqnx
and indeed is given by a gaussian with arbitrary complex $\sigma$. It is a simple exercise to confirm 
Eq. (\ref{B3}) for power like observables, e.g., by calculating the generating function in both representations.
For $Re\; \sigma > 0$ the contour $\Gamma_z$ can be rotated into the real axis and Eq. (\ref{B1}) established.
However (\ref{Pxy}) is more general than the original solution \cite{AY} since it provides the positive representation for arbitrary complex $a$, or equivalently $\sigma \in C$.  For some $\sigma$, for example $\sigma\in R, \sigma < 0$,
the contour $\Gamma_z$ cannot be rotated back into the real axis. Then Eq. (\ref{Pxy}) gives the positive and normalizable representation for the averages along the allowed $\Gamma_z$, or in another words,
for the analytic continuation of the divergent, along the real axis, expressions. 

The aim of this paper is to generalize (\ref{Pxy}) to $N \rightarrow \infty$ variables and apply it to quantum mechanical cases of interest taking $a$ to be purely imaginary.
\section{Many variables}
For the action we take $N$ copies of (\ref{Pzz})  and add the nearest neighbour couplings, with periodic boundary conditions in $z_i$ and $\bz_i$: $z_{N+1}=z_1, \bz_{N+1}=\bz_1, z_{0}=z_N, \bz_{0}=\bz_N$, $a, c \in C$, $b \in R$,
\eqn
S_N(z,\bz)= \sum_{i=1}^N
 a \bz_i^2 + 2 b  \bz_i z_i + 2 c \bz_i z_{i+1} + 2 c ^* z_i \bz_{i+1} + a^* z_i^2 . \label{SN} 
\eqnx
The complex density $\rho(z)$ results from integrating $P_N(z,\bz)$ over all $\bz$ variables
\eqn
\rho(z)=\int \prod_{i=1}^{N}d\bz_i P(z,\bz)=\left(\frac{i}{2}\right)^N\int \prod_{i=1}^{N}d\bz_i \exp{\left(-S_N(z,\bz)\right)} .\nn
\eqnx
The integration is elementary and one obtains for the effective action 
\eqn
S_N^{\rho}\equiv -\log{\left\{ \left(\frac{-4a}{\pi}\right)^{\frac{N}{2}}\rho(\{z\})\right\}}=\sum_{i=1}^N \frac{B}{2a}   
\left(   z_i^2 + 2 \frac{2 b (c+ c^*)}{B} z_i z_{i+1} + z_{i+1}^2 \right) \nn\\
+ \frac{2 c c^*}{a} \left(z_{i-1} z_{i+1}-z_i^2\right),\;\;\;\;\;\; B=b^2+(c+c^*)^2-|a|^2.\;\;\;\;\; \label{Sr2}
\eqnx
If we set  $c$ to be real and require
\eqn
2c=2\gamma=-b + |a|, \label{gam}
\eqnx
 the effective action simplifies to
\eqn
-S_N^{\rho}(z)= {\cal A} \sum_{i=1}^N    \left(   z_i -  z_{i+1} \right)^2
- r\left(z_{i-1}-z_{i+1}\right)^2,\;\;\;\;{\cal A}=\frac{b(b-|a|)}{a} ,\;\;r=\frac{b-|a|}{4b}.\;\;\;\;\label{Srho}
\label{rhN}
\eqnx
This is reminiscent of the discretized Feynman action for a free particle. The second term however, even though
similar to the first one, requires further attention and will be discussed shortly. 

Leaving this for a moment let us check now the positivity and normalizability
of the corresponding probability density $P_N(x,y)$  on $R^{2N}$. In terms of real and imaginary parts of $z_i$ the action (\ref{SN}) reads  
\eqn
S_N(x,y)=
2 \sum_{i=1}^N  (b+\alpha) x_i^2  + 2\beta x_i y_i +(b-\alpha) y_i^2 + 2\gamma (x_i x_{i+1} + y_i y_{i+1}).\label{SNXY}
\eqnx 
Hence  
\eqn
P_N(x,y)=\exp{\left(-S_N(x,y) \right) } \label{PN}
\eqnx
is obviously positive. With $\gamma$ given by (\ref{gam}), all $2N$ eigenvalues are non-negative -- there are no divergent directions. There is one zero mode associated with the translational invariance, however this is usual and can be dealt with by  standard means.  
\section{The continuum limit}
\subsection{A free particle}
The action (\ref{rhN}) does not agree with the standard, discretized action of a free particle 
\eqn
S^{free}_N=\frac{i m}{2 \hbar\epsilon} \sum_{i=1}^N (z_{i+1}-z_i)^2,  \label{SFd}
\eqnx
except at $r=0$.  To see better the effect of the next-to-nearest ($nnn$) term,  we analyze in detail the large $N$ behaviour of, e.g., the propagator
\eqn
K_N(z_N,z_1)=e^{-{\cal A}(z_N-z_1)^2} I_{N}(z_N,z_1) = e^{-{\cal A}(z_N-z_1)^2}\int dz_2...dz_{N-1} e^{-S_N^{\rho}(z_1,...,z_N)}. \;\;\;\;\label{In}
\eqnx
For simplicity we shall work in the ``exponential accuracy", i.e. ignore all prefactors. They can be dealt with by usual methods and do not affect 
any conclusions drawn here. 
 We also  rescale temporarily all variables $ {\cal A} z_i^2 \rightarrow z_i^2$ to further simplify all expressions.
The integral (\ref{In}) can be calculated recursively, $k=2,3,..,N-1$, 
\eqn
 I^{(N)}_k(z_N,z_1;v,w)=\int d u  I^{(N)}_{k-1}(z_N,z_1;u,v) e^{(u-v)^2-r(u-w)^2},  \nn 
\eqnx
with the initial condition
\eqn
I^{(N)}_1(z_N,z_1,u,v)=\exp{\left((z_1-u)^2-r(z_1-v)^2-r (z_N-u)^2\right)}.\label{i1}
\eqnx
The propagator obtains after $N-2$ steps
\eqn
I_{N}(z_N,z_1)=I^{(N)}_{N-1}(z_N,z_1;u,v)\left |_{ (u,v)->(z_N,z_1)} \right. . \nn 
\eqnx
It is straightforward to derive recursion relations for the exponents of $I^{(N)}_k$. Define
\eqn
W_k(u,v)=\log{I^{(N)}_k(z_N,z_1;u,v)}=a_k u^2 + 2 b_k u v+ c_k v^2 + 2d_k u + 2 e_k v + f_k,\nn
\eqnx
then
\eqn
a_{k+1}=1+c_k +\frac{2 b_k -b_k^2 -1 }{1- r+ a_k },&\;\;\;\;&
b_{k+1}=\frac{r - r b_k}{1-r+a_k},\nn\\
c_{k+1}=1-\frac{1}{1 - r + a_k},&\;\;\;\;&
d_{k+1}=e_k+\frac{d_k -b_k d_k}{1 - r + a_k},\nn\\
e_{k+1}=\frac{-r d_k}{1 - r + a_k},&\;\;\;\;&
f_{k+1}=f_k-\frac{d_k^2 }{1 - r + a_k},\nn
\eqnx
with the initial conditions implied by (\ref{i1})
\eqn
a_1=1-r&,\;\;\;b_1=0,&\;\;\;c_1=-r,\;\;\; \nn\\
d_1=-z_1+r z_N,&\;\;\;e_1=r z_1,& f_1=z_1^2-r z_1^2-r z_N^2.\nn
\eqnx
Results are the following: $W_N(z_N,z_1)$ is quadratic and depends only on the difference 
\eqn
W_N(z_N,z_1) = \sigma_N(r) (z_N-z_1)^2,  \nn
\eqnx
as required by the translational invariance. The coefficient $\sigma_N$ is the ratio of two polynomials
\eqn
\sigma_N(r) = \frac{P_N(r)}{Q_N(r)},  \nn
\eqnx
and can be expanded for large $N$ as
\eqn
\sigma_N(r)=v_0(r) + \frac{v_1(r)}{N} + \frac{v_2(r)}{N^2}+... \label{sigN}
\eqnx
At $r=0$, all coefficients $v_i$ vanish except of $v_1(0)=1$. This is the standard Feynman case without the $nnn$ term, c.f. (\ref{rhN},\ref{SFd}).
For $r\ne 0$ however all $v_i$ do not vanish, in particular $v_0\ne 0$. This precludes existence of the continuum limit
\eqn
N \rightarrow \infty,\;\;\; N\epsilon \;\; fixed ,\label{lim2}
\eqnx
which requires 
\eqn
{\cal A} \sigma_N(r) \rightarrow \;\;const.\;\;\;\;\; 
{\cal A} \sim \frac{1}{\epsilon},        \nn
\eqnx
as follows from (\ref{SFd}).  In principle one might consider  renormalizing the divergent term away -- the possibility which should be
looked at in more detail. However we choose here a simpler solution. Both constraints, namely
\eqn
{\cal A} =\frac{b(b-|a|)}{a} \rightarrow \frac{i m}{2 \hbar\epsilon} , \;\; and\;\;\; r=\frac{b-|a|}{4b} \rightarrow 0,  \nn
\eqnx 
can be satisfied in the limit  (referred from now on as $\lim_1$)
\eqn
|a|, b \rightarrow \infty, b-|a|=\frac{ m}{2 \hbar\epsilon}=const.\equiv d,\;\;a=-i |a| . \label{lim1}
\eqnx
This completes the construction of the positive representation for the path integral of a free particle directly in the Minkowski time.

All quantum averages can now be obtained by weighting suitable, i.e. complex in general, observables with the positive and normalizable distribution 
(\ref{PN}), and then taking 
 the limit (\ref{lim1}) followed by the continuum limit (\ref{lim2}). Subsequent applications  illustrate how this works in practice.
\subsection{A harmonic oscillator}
Interestingly this case is also covered by the action (\ref{SN},\ref{SNXY}). The only difference lies in the scaling laws imposed during the  first limiting transition (\ref{lim1}). To see this consider the first term in Eq.(\ref{Sr2}), for real $c=\gamma$,
\eqn
D z_i^2 + 2 E z_i z_{i+1} + D z_{i+1}^2,\;\;\; D=\frac{b^2+4\gamma^2-|a|^2}{2 a},\;\;\;\;E=\frac{2b\gamma }{a}.  \nn
\eqnx
Rewrite it as
\eqn
 -E\left( (z_{i+1}-z_i)^2  - \left(\frac{D}{E}+1\right)(z_i^2+z_{i+1}^2)\right), \nn
\eqnx
and compare with an analogous term in the discretization of the Minkowski action of a harmonic oscillator
\eqn
\frac{i m }{2 \hbar \epsilon}\left( (x_1-x_2)^2 - \frac{\omega^2 \epsilon^2}{2}(x_1^2+x_2^2)\right).  \nn
\eqnx
Therefore, the general positive distribution (\ref{SNXY}) in $2N$ real variables describes a harmonic oscillator
if we identify
\eqn
-\frac{ 2 b \gamma}{a} = \frac{i m}{2 \hbar \epsilon},\;\;\;\;\frac{b^2+4\gamma^2-|a|^2}{4 b \gamma} + 1 = \frac{\omega^2 \epsilon^2}{2}. \label{holi1}
\eqnx
Similarly to the free particle case, the $nnn$ terms will vanish for large $|a|$ and $b$. However the limit has to be taken 
along the trajectory (\ref{holi1}). A possible parametrization in terms of one independent variable $\nu$, is
\eqn
\;\;\;a=-i |a|,\;\;\;\;b=\frac{\mu}{\nu},\;\;\; |a|= \frac{\mu}{\nu} \zeta(\nu,\rho),\;\;\; 2\gamma=-\mu \zeta(\nu,\rho),   \label{hotraj}
\eqnx
where
\eqn
\zeta(\nu,\rho)= \frac{\sqrt{1-2\nu^2\rho+\nu^2\rho^2}-\nu(1-\rho)}{1-\nu^2},   \nn
\eqnx
and $\mu$ and $\rho$ depend on $N$ and parameters of the harmonic oscillator in the continuum
\eqn
\rho=\frac{\omega^2 T^2}{2 (N-1)^2},\;\;\;
\mu=\frac{m (N-1)}{2\hbar T}.   \nn
\eqnx
Vanishing of the $nnn$ term is achieved by taking $\nu\rightarrow 0$. 

This is the main modification compared to the free
particle case. With the first limit taken along the trajectory (\ref{hotraj}) the action (\ref{SNXY}) provides a positive representation for Minkowski path integral of a one-dimensional harmonic oscillator. 

However now one eigenvalue of (\ref{SNXY}) becomes "weakly negative" and the procedure requires additional care.
This is the familiar zero eigenvalue which for general $\gamma$ and imaginary $a$ reads 
\eqn
\lambda_0=2(b-|a|+2\gamma),  \nn 
\eqnx
with the corresponding eigenvector having all equal components. In the free particle case (\ref{gam}) $\lambda_0=0$ reflecting the
translational symmetry. Along the new trajectory (\ref{hotraj}) however, $\lambda_0$ does not vanish and is negative. Moreover, after the first limit
\eqn
\lim_{\nu\rightarrow 0} \lambda_0= -\frac{m\omega^2 T}{4\hbar (N-1)},  \nn
\eqnx
and tends to zero with $N\rightarrow\infty$. The eigenvector remains the same for arbitrary $\gamma$  and becomes the true zero mode in the continuum limit. That is why the mode was called "weakly negative". Therefore one can treat it similarly to the usual zero modes, e.g. fix it. 
 In fact, a negative mode
is simpler than the zero mode since moments of divergent distributions can be defined by the analytic continuation which provides a regularization of the divergent integral. Both ways do not affect the continuum limit as will be seen in the following applications.


\section{Applications}
\subsection{A free particle}
First, we shall calculate the free propagator integrating explicitly the new representation (\ref{SNXY}).
The discretized kernel (\ref{In}) reads
\eqn
K_N(z_N,z_1)=e^{-{\cal A} (z_1-z_N)^2}\int d\bz_1\prod_{j=2}^{N-1} dx_j dy_j d\bz_N \exp{\left(-X^T M X\right)}. \label{KN}
\eqnx
The first factor takes away an additional contribution hidden in $S_N$ (\ref{SNXY}) due to the periodic boundary conditions as explicitly seen in (\ref{rhN}).
Since $z_1$ and $z_N$ are fixed, the first and the last integrals have to be done over $\bz_1$ and $\bz_N$ and not over the real coordinates.
This is part of the construction: only complete traces are represented by integrals of positive distributions over the real variables, 
while deriving quantum amplitudes at fixed end-point requires integration over the corresponding complex, barred variables.
Consequently $X$ is the vector of all variables, $X^T=(z_1,\bz_1,x_2,y_2,x_3...,y_{N-1},z_N,\bz_N)$, and $M$ is the matrix of (\ref{SNXY}) in this mixed representation.
Gaussian integration is simple and one obtains up to a prefactor
\eqn
K_N(z_N,z_1)\sim\exp{\left( \sigma_N(a,b)(z_N-z_1)^2\right)}, \label{sigN}
\eqnx
with $\sigma_N(a,b)$ given in Table 1 for few values of $N$, and  
\eqn
a=-i|a|,\;\;|a|=b-d, d=\frac{m}{2\hbar\epsilon}.  \nn
\eqnx

\begin{table}
\begin{center}
\begin{tabular}[c]{ c  c  c}
\hline\hline
  $N$ & $\sigma_N(-i(b-d),b)$ &  $\lim_1$ \\
  \hline\hline
 5 &  $\frac{ i d(16 b^2 +28 b d -19 d^2)}{8(8b-3d)(b-d)}$& $i\frac{   d}{4}$ \\
 8 & $\frac{ i  d(16 b^4+40b^3 d -70 b^2 d^2 +23 b d^3-d^4)}{(b-d)(112b^3-120 b^2 d +30b d^2 -d^3)}$& $i\frac{  d}{7} $\\
 11 & $\frac{i d (1024 b^5 +3328 b^4 d-9472 b^3 d^2 +6832 b^2 d^3 -1700 b d^4+109 d^5)}{(b-d)(1280b^4-2304 b^3 d+1344 b^2 d^2-280 b d^3+15 d^4)}$&  $i\frac{ d}{10}$\\
 \hline\hline
\end{tabular}
\end{center}
\caption{The slope of the free propagator (\ref{sigN}) and its limiting value for few discretizations.}
\end{table}

Results after the first limit (\ref{lim1}) are given in the third column. Indeed, as discussed in Sect.3, the $v_0$ term (c.f. (\ref{sigN} )) does not survive 
and the limiting $\sigma_N$ has the appropriate large $N$ behaviour 
\eqn
\lim_{b\rightarrow\infty}\sigma_N(-i (b-d),b) = \frac{i d}{N-1},  \nn
\eqnx
which assures the correct  and well known form
\eqn
\lim_{N\rightarrow\infty} K_N\sim\exp{\left(\frac{i m}{2 \hbar}\frac{(z_N-z_1)^2}{T}\right)}.  \nn
\eqnx
This can be analytically continued to the real axes.

As a second example we calculate the average $\la x^2(t) \ra$ with the new representation. Physically this is the dispersion of a Minkowski path  of a free particle at time $t$. The particle is constrained to start from, and return to, the origin after time T. The continuum result, 
\eqn
\la x^2(t) \ra =\frac{\int dx K(0, x;T-t) x^2 K(x,0;t)}{K(0,0;T)} = \frac{i\hbar }{ m } \frac{t(T-t)}{T},
\label{x2av}
\eqnx
is purely imaginary and shows the famous statistical broadening of quantum paths as we move away from the fixed initial/final end points.

In our case this is again covered by (\ref{KN}), with $M$ replaced by its reduction $R$ which does not involve $z_1$ and $z_N$.
 $K_N(0,0)\equiv Z$ provides the normalization. Appropriate average reads
\eqn
\left.  \la  z_k^2 \ra \right|_{z_1=z_N=0} = 
\int d\bz_1 dx_2 dy_2...dy_{N-1} d\bz_N (x_k+ i y_k)^2 \exp{\left(- X^T R X  \right)}/Z\nn\\
= \frac{1}{2} \left( R^{-1}_{2k-2,2k-2} + i (R^{-1}_{2k-2,2k-1}+R^{-1}_{2k-1,2k-2}) - R^{-1}_{2k-1,2k-1}\right),  \;\;\;\;\;\;\label{avP}
\eqnx
and can be easily calculated. After the first limit (\ref{lim1})  it simplifies to
\eqn
\lim_1 \la  z_k^2  \ra =\frac{i}{2 d} \frac{(k-1)(N-k)}{N-1} \stackrel{N\rightarrow\infty}{\longrightarrow} \frac{i \hbar }{m} \frac{t(T-t)}{T}, \nn
\eqnx
which is just the discretized version of (\ref{x2av}) , since 
\eqn
(N-1)\epsilon=T,\;\;\;(k-1)\epsilon=t.  \nn
\eqnx
Again the weight is not entirely positive because of integration over two complex (but  $2N-4$ real) variables.  As said above this is the consequence of the
 zero mode and how it was fixed. It remains to be seen if other ways of dealing with translational symmetry could change that. 
  
 The next applications is free of this problem.
\subsection{A harmonic oscillator}
There is no zero mode here, therefore we define now the average over all periodic trajectories, 
\eqn
\la x^2(T) \ra =\la x^2(0) \ra &=& \frac{\int dx  x^2 K(x, x;T)}{\int dx  K(x, x;T)} , \nn
\eqnx
which measures the width of a periodic Minkowski trajectory with  the length $T$. 
This is the different observable than was considered in the free particle case. 
With
\eqn
K(x_b,x_a;T)&\sim& \exp{\left\{ \frac{i}{\hbar}\frac{m\omega}{2\sin \omega T}
\left( (x_a^2+x_b^2)\cos{\omega T} - 2 x_a x_b\right) \right\}}, \nn 
\eqnx
one easily obtains
\eqn
\la x^2(T) \ra &=& -\frac{i\hbar T}{ 4 m} \frac{\cot{\frac{\omega T}{2}}}{\frac{\omega T}{2}}.  \label{HOav}
\eqnx
In our framework, and upon the discretization, this is given by the straightforward average over 
the positive distribution (\ref{SNXY}) of $2N$  real variables $X^T=(x_1,y_1,...,x_N,y_N)$
\eqn
 \la  z_1^2 \ra = \frac{1}{Z}
\int \prod_{j=1}^N dx_j dy_j  (x_1+i y_1)^2 \exp{\left\{- X^T M X   \right\}}.\label{newav}
\eqnx
Gaussian average is again given by the same combination of matrix elements as in (\ref{avP}) but with the original matrix $M$. In particular  $\la  z_k^2 \ra$ is independent of $k$ due to the invariance under time shifts.

The explicit expression for (\ref{newav}) in terms of $a, b$ and oscillator parameters is somewhat messy. However upon taking the first limit
 along the trajectory (\ref{hotraj}) it simplifies to
\eqn
\lim_{\nu\rightarrow 0} \la  z_1^2 \ra = -\frac{i\hbar T}{m} \frac{P_N(\omega T /2)}{Q_N(\omega T /2)}.  \label{HOlim1}
\eqnx
The first few polynomials $P_N(x)$ and $Q_N(x)$, $x=\omega T/2$, are listed in Table 2. They gradually build up $\cot(x)/4x$ with increasing $N$, c.f. Fig.1, and one readily recovers the continuum result (\ref{HOav}) at $N\rightarrow\infty$.

\begin{table}
\begin{center}
\begin{tabular}[c]{ c  c  }
\hline\hline
  $N$ & $P_N(x)/Q_N(x)$  \\
  \hline\hline
 5 &  $\frac{(x^2-2x-4)(x^2+2x-4) }{x^2(x^4-20x+80)}$ \\
 8 & $\frac{7(128 x^8-12 544x^6+384 160 x^4-3 764 768 x^2+5 764 801) }{32 x^2 (x^2-49)(2x^2-49)(8x^4-392 x^2+2401)}$\\
 11 & $\frac{5(x^5-5x^4-100x^3+375x^2+1875x-3125)(x^5+5x^4-100x^3-375x^2+1875x+3125)}{2x^2(x^{10}-275x^8+27500x^6-1203125x^4+21484375x^2-107421875)}$\\
 \hline\hline
\end{tabular}
\end{center}
\caption{Dispersion of a Minkowski trajectory calculated from the positive representation (\ref{newav}) for few discretizations.}
\label{tab2}
\end{table}  

\begin{figure}[h]
\begin{center}
\includegraphics[width=9cm]{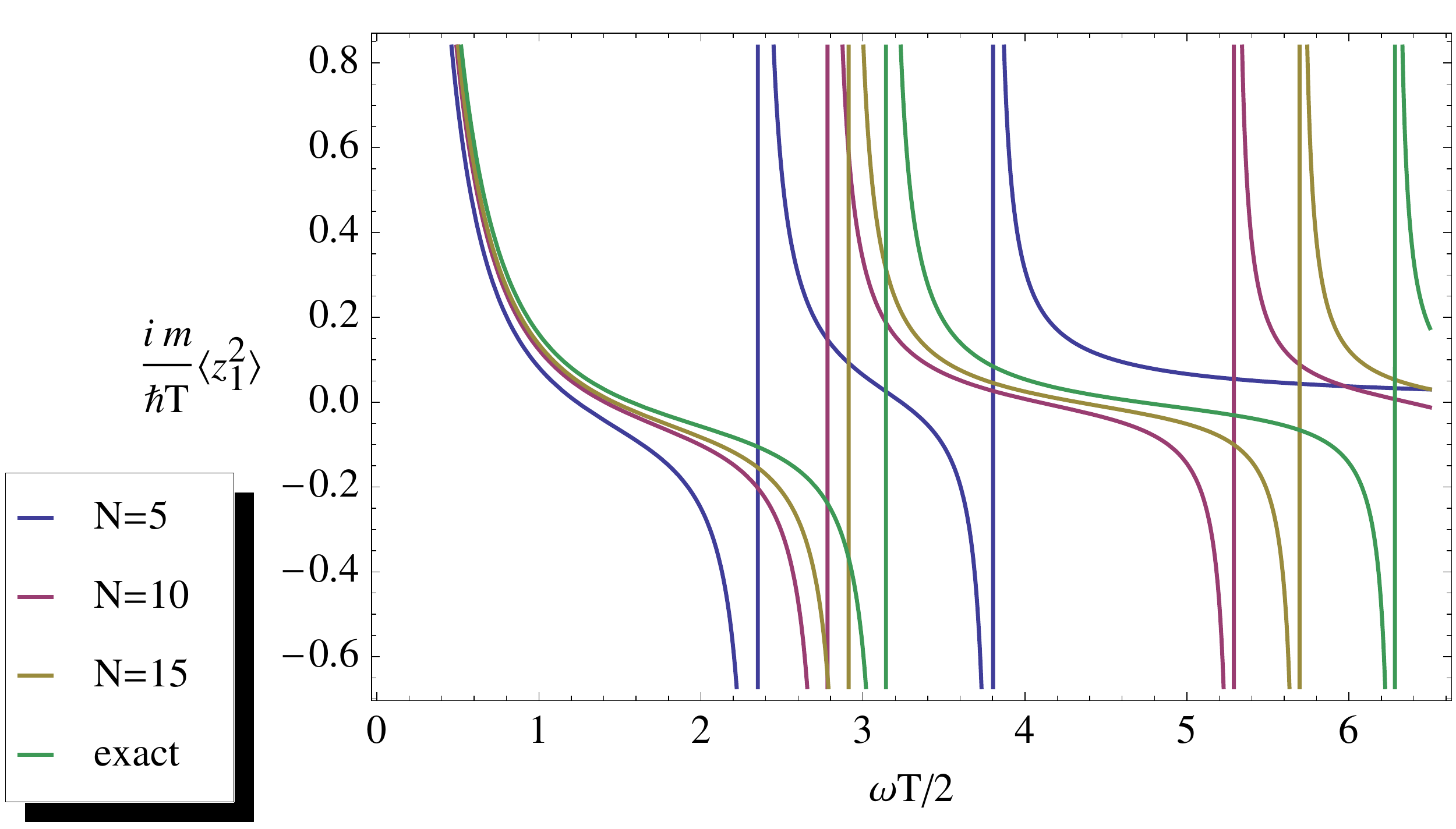}
\end{center}
\vskip-4mm \caption{Convergence of (\ref{HOlim1}) to (\ref{HOav}).} \label{fig:f1}
\end{figure}

The transition from (\ref{HOlim1}) to (\ref{HOav}) is of course well known since the classic papers by Feynman. The novel element here is that (\ref{HOlim1}) was obtained as the probabilistic average of a suitable (i.e. complex) observable over the positive distribution (\ref{PN}).

On the other hand, at finite $N$, the action (\ref{SNXY}) has one negative mode as discussed in Sect 3. Nevertheless the inverse matrix exists meaning that the divergent integral over the negative mode is defined by the analytic continuation. This analytic continuation provides a regularization of the divergence and leads finally to the correct result. Moreover,
by applying the original trick \cite{Ja} a second time, and to the negative mode only, one could construct the positive and normalizable distribution which would allow for statistical calculation of the above and other averages.

\subsection{Charged particle in a constant magnetic field}
This is again the textbook problem in elementary path integrals. It is also the simplest example where Wick rotation does not render the positive Boltzmann factor.
Since the action is again quadratic, it should be possible to construct the corresponding positive density $P_N(x,y)$ similarly to the previous examples. 
Here we shall follow the simpler approach. It is known since the time of Landau that the problem can be reduced to that of a shifted harmonic oscillator.
To use this observation we need to establish the Landau reduction on the level of Feynman propagators. Begin with the phase space path integral 
\eqn
K^B(\vec{x}_b,\vec{x}_a,T)=\int {\cal D} \vec{p}(t) {\cal D} \vec{x}(t) 
 \exp{  \left\{ \frac{i}{\hbar} \left(  \vec{p}.\dot{\vec{x}} - H(\vec{p},\vec{x})  \right)  \right\}}. \label{phsp}
\eqnx
In the gauge used by Landau ($\vec{A}=B(0,x,0)$) the Hamiltonian reads
\eqn
H =\frac{1}{2m} p_x^2 + \frac{1}{2m} \left(p_y - \frac{e B}{c} x\right)^2,  \nn
\eqnx
and one readily obtains  from (\ref{phsp}) , $O=c p_y / B e $,
\eqn
K^B(\vec{x}_b,\vec{x}_a;T)
=\int d O \exp\left\{\frac{i}{\hbar} m\omega O (y_b-y_a)\right\} K_O^{HO}(x_b,x_a;T),\label{fto} 
\eqnx
where $K_O^{HO}(x_b,x_a;T)$ is the kernel for the one dimensional (in $x$) harmonic oscillator located at $x_0=O$. The integral is again gaussian and is 
saturated by the classical position of 
the center of oscillations
\eqn
 O_x=\frac{1}{2}(x_a+x_b) + \frac{1}{2}\cot{\frac{\omega T}{2}}(y_b-y_a).  \label{sad} \nn
\eqnx
Consequently the propagator reads
\eqn
K^B(\vec{x}_b,\vec{x}_a;T)\sim\exp\left\{\frac{i}{\hbar} m\omega O_x (y_b-y_a)\right\} K_{O_x}^{HO}(x_b,x_a;T).  \nn
\eqnx
This (a) corresponds exactly to the Landau solution of the Schr\"{o}dinger equation by separation of variables and (b) after a simple algebra reproduces the
Feynman result in the gauge employed by Landau
\eqn
K_{LG}\sim\exp{ \left\{ \frac{i m}{2 \hbar}  \left(  \frac{\omega}{2} \cot{\frac{\omega T}{2}}\left((x_b-x_a)^2+(y_b-y_a)^2\right) + \omega (  x_a + x_b) (y_b-y_a)  \right)    \right\} }.  \nn
\label{KLG}
\eqnx

Now the reduction (\ref{fto}) can be used to extend our positive representation (\ref{PN} ) also to the case of an external magnetic field. 
Take as an example the average position of a quantum particle at time $0 < t < T $ assuming that at $t=0$ and $t=T$ it was at $\vec{x}_a$ and $\vec{x}_b$ respectively
\eqn
\la \vec{x} \ra_B = \int d^2 x K(\vec{x}_b,\vec{x};T-t) \vec{x} K(\vec{x},\vec{x}_a;t) / K(\vec{x}_b,\vec{x}_a;T) = x^{cl}_{x_a,x_b,T} (t). \label{MFav}   
\eqnx
Since the problem is gaussian the well known, gauge invariant, answer is just the classical trajectory which satisfies above conditions. To see how our representation works in this case
one can use (\ref{fto}) to rewrite (\ref{MFav}) as harmonic oscillator averages
\eqn
\la x(t) \ra_{B}  = \la x(t) \ra_{O=O_x},  \label{rav} \\
\la y(t) \ra_{B}  = \la y(t) \ra_{O=O_y}. \nn
\eqnx
The second line is derived in yet another gauge where the magnetic field problem reduces to the oscillator along the $y$ direction with the analogous classical expression
for the center of $y$ oscillations. 

To complete the construction we only need to extend the positive density (\ref{PN}) such that it describes a shifted harmonic oscillator. This is done by simply adding
 linear terms to the action 
\eqn
S_N(z,\bz) \rightarrow S_N(z,\bz)+\sum_i\; e^* z_i + e \bz_i \label{shft}
\eqnx
or by just shifting $z \rightarrow z_i - z_c$ and $\bz \rightarrow \bz_i - z_c^*$. The new density $P_N$ remains positive and normalizable as before.

Calculation of the appropriate averages in the new representation is now a simple exercise and proceeds analogously to previous applications, e.g. 
(\ref{x2av}). To avoid a confusion with the primordial  cartesian coordinates $x$ and $y$ in (\ref{rav}), we have renamed the real and imaginary parts 
of their complex extensions $z_k$, i.e. $z_k=u_k+i v_k,\;\;\bz_k=u_k-i v_k$. Since the end-points are again fixed the averages are taken over $2N-4$
``positive" variables $u_i,v_i$ and two complex $\bz_1$ and $\bz_N$. Compared to (\ref{KN}) there is an additional source term in the action caused by the shift
(\ref{shft}). The final result obtains after taking the scaling limit ($\lim_1$) defined in (\ref{hotraj}) followed by the usual continuum limit.
\eqn
\la x(t) \ra = \lim_{N\rightarrow\infty} \lim_{\nu\rightarrow 0} \la z_k \ra=\lim_{N\rightarrow\infty} \lim_{\nu\rightarrow 0} \la u_k + i v_k \ra_{P_N(\bz_1,u's,v's,\bz_N)}. \nn
\eqnx
In Fig.2 a sample of averages, after taking the first limit, is shown and compared with the two corresponding classical trajectories, which differ by the choice of the $\omega T$. 
Convergence with $N$ is satisfactory and not surprising. The main point, however, is that the averages are calculated over the new, positive in the fully inclusive case, distribution and they converge in the first limit  to the standard Feynman discretization. 
\begin{figure}[h]
\begin{center}
\includegraphics[width=9cm]{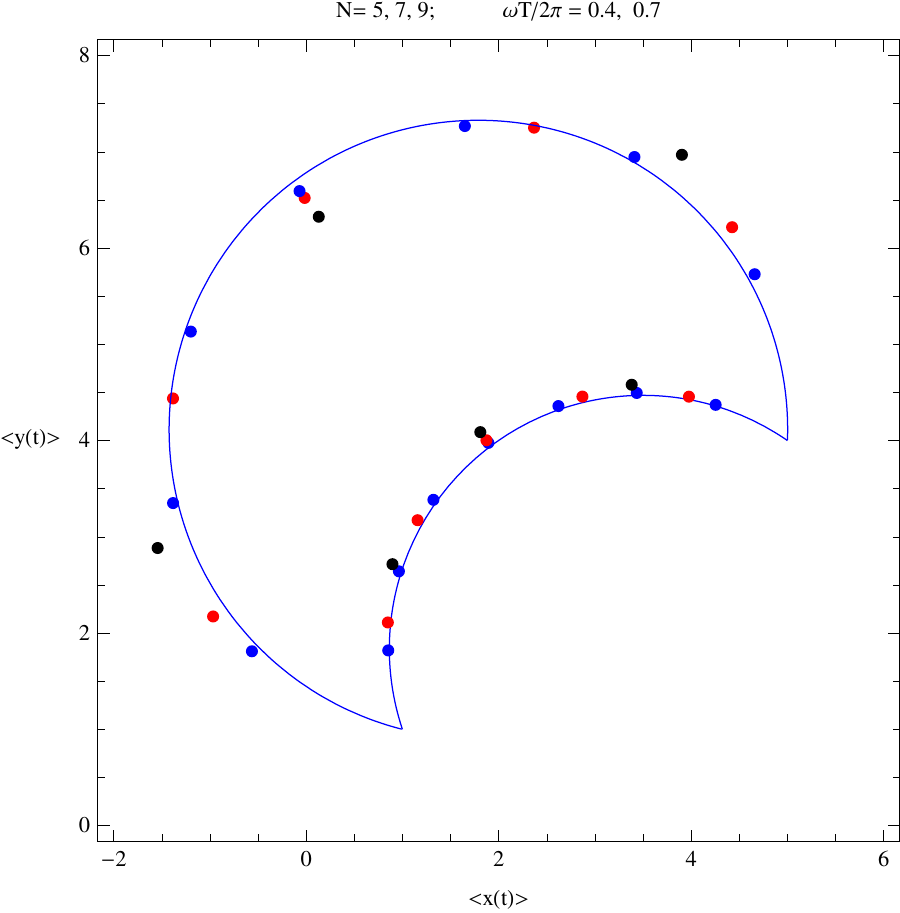}
\end{center}
\vskip-4mm \caption{Two classical trajectories of a charged particle in a constant magnetic field (solid lines). Points represent the first limit of averages (\ref{rav}) calculated with shifted positive density (\ref{PN},\ref{shft}) for finer and finer discretizations.} \label{fig:f2}
\end{figure}
\section{Summary and conclusions}
Problems with complex solutions of the Langevin equations can be avoided by the direct construction of pairs of corresponding complex and positive densities, 
without any reference to complex stochastic processes or Fokker-Planck equations.
This was recently done in Ref.\cite{Ja} for the gaussian model and for its simple quartic modification. As a byproduct the well known solution of the gaussian model
was generalized, thereby providing a positive representation for an arbitrary complex dispersion parameter. In particular it works also for the purely imaginary slope.

In this article the gaussian solution is generalized to many variables and used to construct the positive representation 
for gaussian path integrals directly in the Minkowski time. For the infinite number of degrees of freedom existence of the continuum limit 
is not trivial and is discussed in some details. In particular the couplings appearing in the new representation have to be tuned in a well defined way 
to assure the existence of the continuum limit.

The procedure is then successfully applied to the three textbook quantum mechanical problems: 
a free particle, a harmonic oscillator and a particle in a constant, external magnetic field. The latter is the simplest prototype of a Wilson loop and is known
for its lack of a positive weight after the Wick rotation. 

Many questions remain open, even in the context of above simple cases. For example, how fast is the first limit achieved in practice, how this depends on $N$, is there a more optimal way to
combine the first limit with the continuum limit, etc.

Obviously one would like to generalize the present scheme to nonlinear systems. Related with this is a mathematical problem to what extent can the sum rule (\ref{Pro}), together with positivity and normalizability conditions,
determine $P$ from a complex weight $\rho$. The quartic example solved in Ref.\cite{Ja} shows, that the new structure is not necessarily restricted to the gaussian case. But more systematic study of this question is needed.

A host of further problems and applications suggests itself: generalization to compact integrals, nonlinear and nonabelian couplings, fermionic integrals, as well as extensions to the field theory are only few examples. 
We are looking forward to study some of them.

Finally, an intriguing analogy may be enjoyed.  
Basically the positivity is achieved by duplicating the number of variables.
In these variables,  Minkowski weights become positive as long as boundary conditions for Feynman paths are not specified, i.e. when only traces 
of evolution operators (and/or their moments) are required. Moreover, path integrals in above variables involve a new limiting transition, which may lead, via the saddle point mechanism, to the dominance of a concrete class of trajectories. 
All this resembles to some extent the celebrated history of hidden variables. At the same time we strongly emphasize that none of the sacred principles of quantum mechanics is violated.
The standard, complex quantum amplitudes emerge upon suitable integrations over half of above variables with the usual fixed boundary conditions. Therefore the quantum interference is not violated in any way. Similarly, even though some couplings between new variables indeed have to tend to infinity in the first limit, there are others which remain constant and are in fact $O(1/\hbar)$, hence they drive the usual quantum fluctuations of a system. 

Interestingly Ref. \cite{We} concludes with similar considerations, which however are more speculative due to the lack of the continuum limit analysis.
It will be very interesting to study how the existence of the latter restricts some scenarios  mentioned there.

More generally, it remains to be seen if the new structure exposed in this article is of practical interest only, or if it is more generic.  

\newpage

\noindent {\em Acknowledgements} 

\vspace*{.5cm}

\noindent  I would like to thank Owe Philipsen for the discussion.


\begin{thebibliography}{99}
\bibitem{PW} G. Parisi and Y. Wu, {\em Perturbation theory without gauge fixing}, Scientia Sinica, {\bf 24} (1981) 483.
\bibitem{DH} P. Damgaard and H. H\"{u}uffel {\em Stochastic quantization}, Phys. Rep. {\bf 152} (1987)227.
\bibitem{P} G. Parisi, {\em On complex probabilities}  Phys. Lett. {\bf B131} (1983) 393.
\bibitem{Kl} J. R. Klauder, {\em Coherent-state Langevin equations for canonical quantum systems with applications to the quantized Hall effect} Phys. Rev. {\bf A29} (1984) 2036.
\bibitem{S1} E. Seiler, D. Sexty, I.-O. Stamatescu, {\em  Gauge cooling in complex Langevin for QCD with heavy quarks},
Phys. Lett. B723 (2013) 213. 
\bibitem{S2} G. Aarts et al.,{\em The phase diagram of heavy dense QCD with complex Langevin simulations} Acta Phys.Polon. Supp. {\bf 8} (2015) 2, 405.
\bibitem{AY} J. Ambjorn and S. -K. Yang, {\em Numerical problems in applying the Langevin equation to complex effective actions},
Phys. Lett. {\bf B165} (1985) 140.
\bibitem{AFP} J. Ambjorn, M. Flensburg and C. Peterson, {\em The complex Langevin equations and Monte Carlo simulations of actions with static charges},
Phys. Lett. {\bf B275}[FS17] (1986) 375.
\bibitem{HW} R. W. Haymaker and J. Wosiek, {\em Complex Langevin simulations of non-Abelian integrals}, Phys. Rev. {\bf D37} (1988) 969.
\bibitem{Ph} J. Glesaaen, M. Neuman, O Philipsen, {\em Heavy dense QCD from a 3d effective Lattice theory}, arXiv:1511.00967[hep-lat].
\bibitem{Bl} J.~Bloch, J.~Mahr and S.~Schmalzbauer, ``Complex Langevin in low-dimensional QCD: the good and the
not-so-good'', arXiv:1508.05252 [hep-lat].
\bibitem{S3}  G. Aarts, E. Seiler, I.-O. Stamatescu, {\em The Complex Langevin method: When can it be trusted?}, Phys. Rev. D81 (2010) 054508. 
\bibitem{S4}  G. Aarts, F. A. James, E. Seiler, I.-O. Stamatescu, {\em Complex Langevin: Etiology and Diagnostics of its Main Problem}, Eur. Phys. J. C71 (2011) 1756. 
\bibitem{Ja}  J. Wosiek, {\em Beyond complex Langevin equations I: two simple examples}, arXiv:1511.09083[hep-lat],
\bibitem{We} D. Weingarten, {\em Complex probabilities on $R^N$ as real probabilities on $C^N$ and application to path integrals}, Phys. Rev. Lett. {\em 89} (2002)  240201-1.
\bibitem{Sa} L. L. Salcedo, {\em Representation of complex probabilities}, J. Math. Phys. (N.Y.) {\bf 38} (1997) 1710.
\end{thebibliography}
\end{document}